# Electrons and holes in planar systems and condensation by scattering of electrons-holes pairs in high-TC materials

Y. Ben-Aryeh

*Physics Department Technion-Israel Institute of Technology, Haifa, 32000, Israel*

*E-mail address*: phr65yb@physics.technion.ac.il   (Y.Ben-Aryeh)

**Abstract**

Quantum field-theory is developed for treating electrons and holes in planar systems. Non-unitary representations of Dirac equation in the plane are developed. These equations can be used for treating holes-electron pairing in high TC materials where in a non-relativistic approximation we have a relatively small energy band gap. It is shown that while holes or electrons satisfy separately Pauli-exclusion principle, with anti-commutation relations, the holes-electron pairs follow boson commutation relations. The distribution for electrons-holes pairs in planar systems is calculated for the case in which scattering effects between different pairs can be neglected. It is shown that for high surface density in which many pairs are included in an area with wave length dimension, Bose-Einstein condensation can occur. The ground state of the Bose condensation is calculated and the calculations are compared with those of the BCS theory. The physical implications to high TC currents in planar systems are discussed. The present model is related to the chemistry of high-TC materials.

# 1 Introduction

In some works it has been suggested that certain phenomena of high-temperature superconductivity are related to holes-electrons pairing (see e.g. [1-6]). Since high-temperature superconductivity is of utmost importance for new technology (see e.g. [7]), we would like to check this idea, in the present work, by using a quantum field approach which is different from the usual ones. In the present study we are interested in quantum systems which are described in the plane. Such assumption is approximately correct when the forces are mainly in the planar directions and the forces in the z direction can be neglected. As is well known quantum Hall effects are related to such planar systems (see e.g. [8-10]). In a previous work [11] I have developed a quantum field theory by using a nonrelativistic approximation and the results become consistent with the Schrodinger type equations for quantum Hall effects. Such systems treat, however, only electrons where relativistic effects can appear only as small perturbations. In the present work we are interested in quantum field theory in which both electrons and holes are treated on the same footing. In the literature there are various discussions on the spin-statistics theorem and the question arises if this theorem follows from quantum field theory, like those of the Dirac equation [12-13]. I adopt here such approach, so that for treating planar or semi-planar systems it will be useful to start with a variation of Dirac equation in the plane which will be based on the SO(2,1) group instead of the usual SO(3,1) group. We develop such equation with a corresponding quantum-field theory. While we develop a rigorous relativistic equation for electrons and holes, the theory can be applied to non-relativistic holes-electrons pairing under certain approximations. In the non-relativistic regimes like those of high TC materials non-relativistic energy band gap $E_{gap}$ is obtained. Also in calculation of the momenta we take into account the effective masses for the electron and hole which are dependent on solid state effects. The main interest in the present article is to show that holes-electrons-pairing can lead to boson interactions and discuss the implications of such interactions to high TC materials. It will be shown in the present paper that high TC properties can be related to electron-holes pairs which satisfy boson commutation relations. It has been claimed in the literature (see e.g. [14]) that there is no Bose-Einstein condensation in the plane. I find, however, by following the present analysis, that under high surface density of the electrons-holes pairs Bose-Einstein condensation will occur, .i.e., when many electrons-holes pairs are included in an area with the pair wavelength dimensions. I calculate the ground Bose state, and the similarity and the difference between the present calculations and those made in the BCS theory are explained. I discuss the implications of the present analysis to high TC currents. I give a short discussion about the relations between the present model and the chemistry of high TC materials.

The paper is organized as follows:

In Section 2, we study the problem of relativistic physics in the plane in which the ordinary Dirac equation, based on SO(3,1) group [15], is reduced to a relativistic equation based on SO(2,1) group [16]. We should notice that there are not any finite dimensional unitary representations of the Lorentz group and the Lorentz group solutions are finite only as non-unitary representations. So, for finite dimensional systems we are forced to take non-unitary representations. We assume that special Lorentz equation can be used for the $x, y, t$ coordinates and it leads to certain new non-unitary representations. The special result which appears in the present formalism is that the wave functions of this relativistic equation are *two-dimensional* in comparison with the wave functions of the ordinary Dirac equation which are *four dimensional*. The normalization relations of these wave functions are based on the properties of non-unitary metric [17-19]. Astonishing point in this analysis is that the spin functions are not included in the present Dirac equations in the plane, and in this respect it is similar to basic equations for quantum Hall effects in the plane which do not include spin functions. The spin functions should be taken into account in relation to spin-statistics theorem [12-13], but within the present approximations, *interactions* with spin functions are neglected.

The quantization of the SO(2,1) relativistic equation in the $x, y, t$ coordinates is made in Section 3 of the present paper by using quantum field methods analogous to those used for the ordinary Dirac equation [20-22] in the $x, y, z, t$ coordinates. The simple approach of having 'holes' with negative energies raises a fundamental problem as the energy is not bounded from below, which cannot be correct. In analogy to the ordinary Dirac equation the solution for this problem is to use a quantum field theory in which we expand the quantum field operators $\hat{\Psi}$ and its conjugate $\hat{\bar{\Psi}}$ into linear combinations of plane waves solutions, and quantize the fields with creation and annihilation operators. In this theory creation operators $\hat{d}^\dagger$ for holes are combined with annihilation operators $\hat{b}$ for electrons, and vice versa. The negative energy solutions for electrons will represent positive energy solutions for holes so that the problem of having negative energy solutions is eliminated. In order to implement this theory we need to assume that the creation and annihilation operators satisfy anti-commutation relations, which have very important consequences for electrons and holes statistics [12-13]. We need also to use the normalization conditions for the wave functions which follow from the properties of the non-unitary metric [17-19]. We find that wave functions for the electrons in quantum hall effects are one dimensional, but we are interested in two-dimensional wave functions for electrons and holes, respectively, under a nonrelativistic approximation.

In Section 4 we use a non-relativistic approximation in which the relativistic energy gap of $2mc^2$ is changed into the small energy band gap $E_{gap}$. I describe a simple solid state model for electrons-holes pairs obtained by the transition of an electron from the filled valence band to the empty conduction band. I describe the Hamiltonian for the creation and annihilation of such pairs. The creation and annihilation operators for electrons-holes pairs satisfy the boson commutation relations although electron and holes satisfy separately the anti-commutation relations. For each pair the electron and hole are attracted to each other by the Coulomb potential, so that a bound state is produced. A bound electron-hole pair can move through the crystal transporting excitation energy but not charge. Such description is correct as long as scattering effects between different pairs can be neglected. The situation becomes quite different when many pairs are included in an area with the pair wavelength dimensions. Under such condition attractive scattering effects occur between different pairs. The distribution of the electrons-holes pairs is calculated in Section 4 showing the condition under which attractive scattering effects become important.

I analyze in Section 5 the possibility that Bose condensation can occur under the condition of high surface density of pairs, due to the attractive interaction between different pairs. The analysis is somewhat similar to that of the BCS theory (see e.g. [23-24]), but becomes different due to the fact that we consider electrons-holes pairs which is a different system from that of two-electron system mediated by phonon interaction in BCS analysis. An equation for the Bose condensation energy gap is obtained which is different from that of BCS.

Although we use in the present system a very simple model for Bose condensation of electrons-holes pairs, I claim that similar effects occur in the complicated high-TC systems. I discuss In Section 6, relations between the High-TC materials and the present "simple" model, under which many pairs are included in an area with wavelength dimensions. In Section 7 the present results and conclusions are summarized.

## 2  The Wave functions for the SO(2,1) Relativistic Equation

We assume that the relativistic equation which is a function of $x, y, t$ coordinates can be given as:

$$\left[ -i\frac{\sigma_0}{c}\frac{\partial}{\partial t} + \sigma_1 \frac{\partial}{\partial x} + \sigma_2 \frac{\partial}{\partial y} + \frac{mc}{\hbar} I \right] \Psi_\pm = 0 \quad . \tag{1}$$

Here $\sigma_1, \sigma_2$ and $\sigma_0 \equiv \sigma_3$ are the Pauli matrices, $m$ the electron mass, $c$ the light velocity, $I$ the unit $2\times 2$ matrix and $\Psi_\pm$ are the two-dimensional wave functions solutions of Eq. (1). This equation is analogous to the Dirac equation in the four $x, y, z, t$ coordinates. Using the definitions

$$\gamma_0 = -i\sigma_0 \quad , \quad \gamma_1 = \sigma_1 \quad , \quad \gamma_2 = \sigma_2 \quad , \quad x_0 = ct \quad , \quad x_1 = x \quad , \quad x_2 = y \quad , \tag{2}$$

Eq. (1) can be written in analogy to the ordinary Dirac equation as

$$\left(\gamma_\mu \frac{\partial}{\partial x_\mu} + \frac{mc}{\hbar} I\right)\Psi_\pm = 0 \quad , \quad \mu = 0, 1, 2 \quad . \tag{3}$$

The mathematical reasons for assuming such equation are as follows:

a) $\sigma_1, \sigma_2$ and $\sigma_3$ satisfy the Clifford algebra which has been essential for deriving the ordinary Dirac equation:

$$\{\sigma_i, \sigma_j\} = \delta_{i,j} \quad , \quad (i, j = 1, 2, 3) \quad , \tag{4}$$

where the curled brackets denote the usual anti-commutation relation, assuming that Pauli two-dimensional matrices replace the four dimensional Dirac $\gamma$ matrices.

b) We define:

$$K_1 = \frac{-i\sigma_2}{2} \quad , \quad K_2 = \frac{i\sigma_1}{2} \quad , \quad K_3 = \frac{\sigma_0}{2} \quad , \tag{5}$$

and then we get the commutation-relations of the SO(2,1) algebra :

$$[K_1, K_2] = -iK_3 \quad , \quad [K_3, K_1] = iK_2 \quad , \quad [K_2, K_3] = iK_1 \quad . \tag{6}$$

Using the definition of Eq. (5) the present relativistic equation can be written as

$$\left[\frac{-2iK_3}{c}\frac{\partial}{\partial t} - 2iK_2\frac{\partial}{\partial x} + 2iK_1\frac{\partial}{\partial y} + \frac{mc}{\hbar} I\right]\Psi = 0 \quad . \tag{7}$$

Eqs. (1), (3) and (7) express the same relativistic equation in different forms. Eq. (1) will be used in the present Section for deriving the wave functions solutions of the relativistic equation and their normalization relations. The quantization of this relativistic equation, according to quantum field theory, will be developed in Section 3. In Section 4 we discuss application of the present theory for electrons-holes pairing where the relativistic gap energy $2mc^2$ is exchanged into the band gap $E_{gap}$, and which under certain conditions can lead to high TC currents.

c) By squaring the operators in front of the $\Psi_\pm$ fields we get:

$$-\frac{1}{c^2}\frac{\partial^2}{\partial t^2}\Psi = -\left(\frac{\partial^2}{\partial x^2}+\frac{\partial^2}{\partial y^2}\right)\Psi + \left(\frac{mc}{\hbar}\right)^2 \Psi \quad , \tag{8}$$

which is the two-dimensional wave equation of the Klein-Gordon form.

d) The solutions of (1) can be given as :

$$\Psi_+ = u(k_x, k_y)\exp\left[i(k_x x + k_y y - \omega t)\right] \quad , \tag{9}$$

$$\Psi_- = v(k_x, k_y)\exp\left[-i(k_x x + k_y y - \omega t)\right] \quad , \tag{10}$$

where $u$ and $v$ are two-dimensional wave functions which are functions of $k_x$ and $k_y$, where $k_x$ and $k_y$ are the components of the wave vector in the $x$ and the $y$ direction, respectively, $\omega$ is the frequency, and the homogeneous equations obtained by substituting (9-10) into (1) have solutions only under the condition

$$\omega^2 = (k_x^2 + k_y^2)c^2 + \left(\frac{mc^2}{\hbar}\right)^2 \quad . \tag{11}$$

Eq. (11) can be transformed into the form

$$E^2 = (p_x^2 + p_y^2)c^2 + (mc^2)^2 \quad , \quad (E = \hbar\omega \quad , \quad p_x = \hbar k_x \quad , \quad p_y = \hbar k_y), \tag{12}$$

so that the relativistic equation (11) or equivalently (12) is obtained.

Let us evaluate in the following analysis the classical solutions of (1), by following the above theoretical scheme. In the rest frame $(k_x = k_y = 0)$ we get

$$\left[-i\frac{\sigma_0}{c}\frac{\partial}{\partial t}\right]\Psi_\pm + \left(\frac{mc}{\hbar}I\right)\Psi_\pm = 0 \quad , \tag{13}$$

and the solutions are

$$\begin{aligned}\Psi_+ &= \exp\left[-i\left(\frac{mc^2}{\hbar}\right)t\right]\begin{pmatrix}1\\0\end{pmatrix} \Rightarrow \hbar\omega = mc^2 \quad ,\\ \Psi_- &= \exp\left[i\left(\frac{mc^2}{\hbar}\right)t\right]\begin{pmatrix}0\\1\end{pmatrix} \Rightarrow \hbar\omega = -mc^2\end{aligned} \tag{14}$$

We encounter already here the problem of obtaining a negative energy solution for the state $\psi_-$, often referred to as 'hole'.

Assuming $k_x \neq 0$, $k_y \neq 0$, then the *un-normalized* solution of $u(k_x, k_y)$ can be given as

$$u(k_x, k_y) = \begin{pmatrix} 1 \\ G_1(k_x, k_y) \end{pmatrix} \quad , \tag{15}$$

so that $\Psi_+$ of (1) is given as

$$\Psi_+ = \begin{pmatrix} 1 \\ G_1(k_x, k_y) \end{pmatrix} \exp\left[i(k_x x + k_y y - \omega t)\right] \quad . \tag{16}$$

For evaluating $G_1(k_x, k_y)$ we use (16) and the relations:

$$-i\frac{\sigma_0}{c}\frac{\partial}{\partial t}\Psi_+ = \begin{pmatrix} -\omega/c & 0 \\ 0 & \omega/c \end{pmatrix}\Psi_+ \quad , \tag{17}$$

$$\left(\sigma_1\frac{\partial}{\partial x} + \sigma_2\frac{\partial}{\partial y}\right)\Psi_+ = \begin{pmatrix} 0 & i(k_x - ik_y) \\ i(k_x + ik_y) & 0 \end{pmatrix}\Psi_+ \equiv H\Psi_+ \quad . \tag{18}$$

We notice that the matrix $H$ defined here is non-Hermitian but satisfy the property [17-19]

$$\eta H \eta^{-1} = H^\dagger \tag{19}$$

with the metric $\eta \equiv \sigma_0$. We verify that

$$\sigma_0 H \sigma_0^{-1} = \begin{pmatrix} 1 & 0 \\ 0 & -1 \end{pmatrix}\begin{pmatrix} 0 & i(k_x - ik_y) \\ i(k_x + ik_y) & 0 \end{pmatrix}\begin{pmatrix} 1 & 0 \\ 0 & -1 \end{pmatrix} = \begin{pmatrix} 0 & -i(k_x - ik_y) \\ -i(k_x + ik_y) & 0 \end{pmatrix} \equiv H^\dagger \tag{20}$$

We will take into account later the use of the metric $\eta \equiv \sigma_0$ for normalization relations.

By substituting (16-18) into (1), and multiplying by $c$, we get:

$$\begin{bmatrix} -(\omega - mc^2/\hbar) & ic(k_x - ik_y) \\ ic(k_x + ik_y) & (\omega + mc^2/\hbar) \end{bmatrix}\begin{pmatrix} 1 \\ G_1 \end{pmatrix} = 0 \quad . \tag{21}$$

The homogeneous equations (21) have solutions only under the condition that the determinant of the coefficients in the square brackets vanishes, which leads to (11)

Separating (21) into its components we get:

$$ic(k_x - ik_y)G_1 = (\omega - mc^2/\hbar) \quad , \tag{22}$$

$$-ic(k_x + ik_y) = (\omega + mc^2/\hbar)G_1 \quad . \tag{23}$$

From (23) we get

$$G_1 = \frac{-i(k_x + ik_y)c}{(\omega + mc^2/\hbar)} = \frac{-i(p_x + ip_y)c}{E + mc^2} \quad . \tag{24}$$

By substituting (24) into (22) we verify the validity of this equation.

For evaluating $\Psi_-$, in the general case, we assume that the *un-normalized* solution of $v(k_x, k_y)$ can be given as

$$v(k_x, k_y) = \begin{pmatrix} G_2(k_x, k_y) \\ 1 \end{pmatrix} \quad , \tag{25}$$

so that $\Psi_-$ of (1) is given as

$$\Psi_- = \begin{pmatrix} G_2(k_x, k_y) \\ 1 \end{pmatrix} \exp\left[-i(k_x x + k_y y - \omega t)\right] \quad . \tag{26}$$

Here, again, $k_x$ and $k_y$ are the components of the wave vector in the $x$ and $y$ directions, $\omega$ is the frequency, and $G_2$ is a function of $k_x, k_y$.

For evaluating $v(k_x, k_y)$ we use the relations:

$$-i\frac{\sigma_0}{c}\frac{\partial}{\partial t}\Psi_- = \begin{pmatrix} \omega/c & 0 \\ 0 & -\omega/c \end{pmatrix}\Psi_- \quad , \tag{27}$$

$$\left(\sigma_1 \frac{\partial}{\partial x} + \sigma_2 \frac{\partial}{\partial y}\right)\Psi_- = \begin{pmatrix} 0 & -i(k_x - ik_y) \\ -i(k_x + ik_y) & 0 \end{pmatrix}\Psi_- \equiv H\Psi_- \quad . \tag{28}$$

Here again the matrix $H$ is non-unitary but satisfy the condition of Eq. (19) with the same metric $\eta \equiv \sigma_0$.

By substituting (26-28) for $\Psi_-$ into (1), and multiplying with $c$, we get:

$$\begin{pmatrix} (\omega + mc^2/\hbar) & -i(k_x - ik_y)c \\ -i(k_x + ik_y)c & -(\omega - mc^2/\hbar) \end{pmatrix} \begin{pmatrix} G_2 \\ 1 \end{pmatrix} = 0 \quad . \tag{29}$$

Here, again, the homogeneous equations (29) have solutions only under the condition that the determinant of the coefficients in the square brackets vanishes, which leads again to (11).

Separating (29) into its components we get:

$$ic(k_x - ik_y) = (\omega + mc^2/\hbar)G_2 \quad , \tag{30}$$

$$-ic(k_x + ik_y)G_2 = (\omega - mc^2/\hbar) \quad . \tag{31}$$

From (30) we get:

$$G_2 = \frac{i(k_x - ik_y)c}{(\omega + mc^2/\hbar)} = \frac{i(p_x - ip_y)c}{(E + mc^2)} \tag{32}$$

By substituting (32) into (31), we verify the validity of this equation.

One should notice that the above wave functions are based on non-unitary representations so that they satisfy special normalization relations given as follows. The normalization relations are obtained between the two-dimensional vectors $u(k_x, k_y)$ and $v(k_x, k_y)$, and $u(k_x, k_y)^\dagger$ and $v(k_x, k_y)^\dagger$, where $u(k_x, k_y)^\dagger = \bar{u}(k_x, k_y)\sigma_0$ and $v(k_x, k_y)^\dagger = \bar{v}(k_x, k_y)\sigma_0$, and where $\bar{u}$ and $\bar{v}$ are the conjugate of $u$ and $v$, respectively. One should notice that $\sigma_0$ has represented a certain metric for the non-unitary representations. Following these definitions we get:

$$\bar{u}(k_x, k_y)\sigma_0 u(k_x, k_y) = (1, \bar{G}_1)\begin{pmatrix} 1 & 0 \\ 0 & -1 \end{pmatrix}\begin{pmatrix} 1 \\ G_1 \end{pmatrix} = 1 - |G_1|^2 \tag{33}$$

The normalized wave function (indicated by the subscript N) is given by

$$u_N = \begin{pmatrix} 1 \\ G_1 \end{pmatrix} / \sqrt{1 - |G_1|^2} \tag{34}$$

By using (15) and (32) we get the relation

$$\bar{u}(k_x, k_y)\sigma_0 v(k_x, k_y) = (1, \bar{G}_1)\begin{pmatrix} 1 & 0 \\ 0 & -1 \end{pmatrix}\begin{pmatrix} G_2 \\ 1 \end{pmatrix} = G_2 - \bar{G}_1 = 0 \tag{35}$$

Normalization of $v(k_x, k_y)$, representing the negative energy solutions, is given by:

$$\bar{v}(k_x, k_y)\sigma_0 v(k_x, k_y) = (\bar{G}_2, 1)\begin{pmatrix} 1 & 0 \\ 0 & -1 \end{pmatrix}\begin{pmatrix} G_2 \\ 1 \end{pmatrix} = -1(1 - |G_2|^2) \tag{36}$$

where $|G_2| = |G_1|$. The normalized $v_N(k_x, k_y)$ function is then given by

$$v_N(k_x, k_y) = \begin{pmatrix} G_2 \\ 1 \end{pmatrix} / \sqrt{1 - |G_2|^2} \tag{37}$$

Summarizing the -normalization relations we have:

$$u_N^\dagger(k_x, k_y) v_N(k_x, k_y) = v_N^\dagger(k_x, k_y) u_N(k_x, k_y) = 0 \tag{38}$$

$$u_N^\dagger(k_x, k_y) u_N(k_x, k_y) = 1 \quad , \quad v_N^\dagger(k_x, k_y) v_N(k_x, k_y) = -1 \tag{39}$$

The negative value for the normalization of $v_N(k_x, k_y)$ leads to a norm which is not positive definite and the solution for this problem is to add to the metric the operator $C$ which is identical to charge conjugation in quantum field theory [17-18] and multiplies the metric $\sigma_0$ of $v_N(k_x, k_y)$ by -1. The ortho-normalization relations of (38-39) will be used in the next Section where we develop the SO(2,1) relativistic quantum field theory in which $\hat{\Psi}$ and $\hat{\bar{\Psi}}$ are field operators, leading to propagating plane waves, which are described with the use of creation and annihilation operators. The problem of negative norm for the state $v_N(k_x, k_y)$ will be solved also there.

## 3 Quantization of the relativistic equation in the plane

In the "simple" approach presented in the previous Section we get the energy $\hbar\omega$ for the state represented by $\Psi_+$, and the energy $-\hbar\omega$ for the state represented by $\Psi_-$, where $\omega$ is the positive value given by (11). This simple approach of having 'holes' with negative energies raises a fundamental problem as the energy is not bounded from below, and it leads to a negative norm which cannot be correct. In analogy to the ordinary Dirac equation the solution for this problem is to use a quantum field theory in which we expand the quantum field operators $\hat{\Psi}$ and conjugate $\hat{\bar{\Psi}}$ into linear combinations of plane wave solutions, by (9-10), with creation and annihilation operators. In this theory creation operators $\hat{d}^\dagger$ for positrons (or equivalently for holes) are combined with annihilation operators $\hat{b}$ for electrons, and vice versa. In order to implement this theory one needs to assume that the creation and annihilation operators will satisfy anti-commutation relations, which have very important consequences for electrons - holes statistics [12-13]. The quantum field theory is developed as follows.

The classical *free field Lagrangian* density of the three-dimensional Dirac equation is given by

$$L = -c\hbar \bar{\Psi} \gamma_\mu \frac{\partial}{\partial x_\mu} \Psi - mc^2 \bar{\Psi}\Psi \quad . \tag{40}$$

The independent fields are considered to be the two components of $\Psi$ and the two components of $\bar{\Psi}$. The Euler-Lagrange equation using independent fields $\bar{\Psi}$ is given as

$$\frac{\partial}{\partial x_\mu}\left(\frac{\partial L}{\partial \bar{\Psi}/\partial x_\mu}\right) - \frac{\partial L}{\partial \bar{\Psi}} = 0 \quad . \tag{41}$$

Since there is no derivative of $\bar{\Psi}$ in (41) we get the simple equation

$$\frac{\partial L}{\partial \bar{\Psi}} = 0 \qquad . \tag{42}$$

By substituting Eq. (40) into Eq. (42) we get:

$$-c\hbar \gamma_\mu \frac{\partial}{\partial x_\mu}\Psi - mc^2\Psi = 0 \Rightarrow \left(\gamma_\mu \frac{\partial}{\partial x_\mu} + \frac{mc}{\hbar}\right)\Psi = 0 \quad . \tag{43}$$

Eq. (43) is equal to the relativistic equation given in (3), indicating that the Lagrangian (40) is the right one. To compute the Hamiltonian density, we start by finding the momenta conjugate to the fields $\Psi$ given as

$$\Pi = \frac{\partial L}{\partial \left(\frac{\partial \Psi}{\partial t}\right)} = -c\hbar \bar{\Psi} \gamma_0 / c \qquad . \tag{44}$$

The Hamiltonian density is then given by

$$H = \Pi \frac{\partial \Psi}{\partial t} - L = -c\hbar \bar{\Psi} \gamma_0 \frac{1}{c}\frac{\partial \Psi}{\partial t} + c\hbar \bar{\Psi} \gamma_0 \frac{\partial}{\partial x_0}\Psi + c\hbar \bar{\Psi} \gamma_1 \frac{\partial \Psi}{\partial x} + c\hbar \bar{\Psi} \gamma_2 \frac{\partial \Psi}{\partial y} \quad , \tag{45}$$

where the first two terms on the right side of Eq. (45) are cancelled so that

$$H = c\hbar \left[\bar{\Psi}\left(\gamma_1 \frac{\partial}{\partial x} + \gamma_2 \frac{\partial}{\partial y} + mc/\hbar\right)\Psi\right] \quad . \tag{46}$$

Using (1-3) we apply the transformation

$$\left(\gamma_1 \frac{\partial}{\partial x} + \gamma_2 \frac{\partial}{\partial y} + mc/\hbar\right)\Psi = \left(\frac{i\sigma_0}{c}\right)\frac{\partial}{\partial t}\Psi \quad , \tag{47}$$

so that (46) can be transformed into the simpler form

$$H = \hbar \left[\bar{\Psi}(i\sigma_0)\frac{\partial}{\partial t}\Psi\right] \quad . \tag{48}$$

In analogy with the quantum field theory of the Dirac equation [21-23] in which $\Psi$ and $\bar{\Psi}$ become field operators we get here for the relativistic equation:

$$\hat{\Psi}(x,y,t) = \iint dk_x dk_y \left(\frac{1}{2\pi}\right)\begin{bmatrix}\hat{b}(k_x,k_y)u_N(k_x,k_y)\exp\left[-i(k_x x + k_y y - \omega t)\right]+\\ \hat{d}^\dagger(k_x,k_y)v_N(k_x,k_y)\exp\left[i(k_x x + k_y y - \omega t)\right]\end{bmatrix}, \tag{49}$$

$$\hat{\bar{\Psi}}(x,y,t) = \iint dk_x dk_y \left(\frac{1}{2\pi}\right)\begin{bmatrix}\hat{b}^\dagger(k_x,k_y)\bar{u}_N(k_x,k_y)\exp\left[i(k_x x + k_y y - \omega t)\right]+\\ \hat{d}(k_x,k_y)\bar{v}_N(k_x,k_y)\exp\left[-i(k_x x + k_y y - \omega t)\right]\end{bmatrix}. \tag{50}$$

We take into account that all operators operating on the electrons commute with all operators operating on the holes. The form of $\hat{\Psi}(x,t)$ and $\overline{\hat{\Psi}}(x,t)$ is adapted here for the plane Dirac equation. The creation and annihilation operators satisfy the anti-commutation relations

$$\{\hat{b},\hat{b}\} = \{\hat{d},\hat{d}\} = \{\hat{b},\hat{d}\} = \{\hat{b},\hat{d}^\dagger\} = \{\hat{d},\hat{b}^\dagger\} = 0 \quad , \tag{51}$$

$$\{\hat{b}(k_x,k_y),\hat{b}^\dagger(k'_x,k'_y)\} = \{\hat{d}(k_x,k_y),\hat{d}^\dagger(k'_x,k'_y)\} = \delta(k_x - k'_x)\delta(k_y - k'_y) \quad . \tag{52}$$

In the present quantum field theory (48) represents Hamiltonian density and the Hamiltonian is obtained by integrating the density of the Hamiltonian in the plane so that

$$H = \iint dxdy\hbar \left[\overline{\hat{\Psi}}(x,y,t)i\sigma_0 \frac{\partial}{\partial t}\hat{\Psi}(x,y,t)\right] = \iint dxdy\hbar \overline{\hat{\Psi}}(x,y,t)\sigma_0 \omega(k_x,k_y)\hat{\Psi}(x,y,t). \tag{53}$$

We notice that in (53) the metric $\sigma_0$ is appearing between $\overline{\hat{\Psi}}(x,y,t)$ and $\hat{\Psi}(x,y,t)$. We use here a special form for quantizing $\hat{\Psi}(x,y,t)$ and $\hat{\Psi}^\dagger(x,y,t)$ in the plane, given by (49-50), and where $u_N(k_x,k_y)$ and $v_N(k_x,k_y)$ satisfy the normalization conditions of (38-39), and where the anti-commutation relations of (51-52) are satisfied. We substitute $\hat{\Psi}(x,y,t)$ and $\overline{\hat{\Psi}}(x,y,t)$ from (49-50) into (53) and simplify further the expression for the Hamiltonian by performing first the integration over the plane coordinates. Such integration leads to the delta function multiplication $\delta(k_x - k'_x)\delta(k_y - k'_y)$ so that by using these delta functions the integration in momentum space is reduced to $\iint dk_x dk_y$ and then we get:

$$H = \iint dk_x dk_y \hbar\omega(k_x,k_y)\left[\hat{b}^\dagger(k_x,k_y)\hat{b}(k_x,k_y) - \hat{d}(k_x,k_y)\hat{d}^\dagger(k_x,k_y)\right] \quad . \tag{54}$$

The negative term in Eq. (54) follows from the normalization conditions of (38-39). Using the anti-commutation relations of (52) we can substitute in (54) the relation

$$\hat{d}(k_x,k_y)\hat{d}^\dagger(k_x,k_y) = -\hat{d}^\dagger(k_x,k_y)\hat{d}(k_x,k_y) + \delta(k_x)\delta(k_y) \quad , \tag{55}$$

and by neglecting the terms with zero momentum we get

$$\mathrm{H} = \iint dk_x dk_y \hbar\omega(k_x,k_y)\left[\hat{b}^\dagger(k_x,k_y)\hat{b}(k_x,k_y) + \hat{d}^\dagger(k_x,k_y)\hat{d}(k_x,k_y)\right] \quad . \tag{56}$$

$H$ is obviously a non-negative operator. We find here that the elimination of the negative terms by the use of (55) is equivalent to the use of charge conjugation operator which is needed to eliminate the negative energy terms [17-18]. Assuming that $n^{\pm}(k_x, k_y)$ are the occupation numbers of electrons and holes with momenta $(k_x, k_y)$ then the energy of the state is given by

$$E = \iint dk_x dk_y \hbar \omega(k_x, k_y) \left[ n^+(k_x, k_y) + n^-(k_x, k_y) \right] \geq 0 \tag{57}$$

and the momentum of the is given by

$$\vec{P} = \iint dk_x dk_y \hbar \left( k_x \hat{x} + k_y \hat{y} \right) \left[ n^+(k_x, k_y) + n^-(k_x, k_y) \right] \tag{58}$$

We should take into account that the canonical quantization rules lead to the following anti-commutation relations:

$$\{\Psi, \Psi\} = \{\Psi^\dagger, \Psi^\dagger\} = 0 \quad , \quad \{\Psi(x, y, t), \Psi^\dagger(x', y', t)\} = \delta(x - x') \delta(y - y') \tag{59}$$

where

$$\Psi^\dagger(x, y, t) = \bar{\Psi}(x, y, t) \sigma_0 \quad . \tag{60}$$

In a similar way to the calculation made for the Hamiltonian, which are based on the normalization relations, we need also here to introduce the metric $\sigma_0$ so that the integration is made between $\Psi^\dagger(x', y', t) = \bar{\Psi}(x', y', t) \sigma_0$ and $\Psi(x, y, t)$.

## 4 Electrons-holes pairing in the plane

It is well known fact from the field of solid state physics that holes-electron pairs can be created and annihilated where the electron-hole pair can, for example, be created by the absorption of a photon and annihilated by recombination and photon emission [25-26]. The relativistic equation in the plane is relevant to electrons-holes pairs but it should be modified so that the electrons-holes excitation energy will not be relativistic. In such theory the energy of $2mc^2$ should be exchanged into the non-relativistic band gap energy $E_{gap}$ of the electron-hole pair which is relatively very small. Also the theory for creation and annihilation of holes-electron pairs in solid state physics might include quite complicated solid state phenomena. In spite of these complications we will show here, by using a simple model, an interesting quantum effect, in which the creation and annihilation operators for electrons-holes pairs will satisfy boson commutation relation, although the creation and annihilation operators for each separate electron or hole will satisfy the anti-commutation relations, given by (51-52). We will show physical effects following from this property obtaining High TC currents in such system.

Let us describe in a simple model the transition of an electron from the filled valence band to the conduction band [25-26] obtained by the absorption of a photon which is producing an electron-hole pair with energy

$$\varepsilon(k) = \frac{\hbar^2 k^2}{2}\left(\frac{1}{m_{el}} + \frac{1}{m_{hole}}\right) + E_{gap} = \frac{\hbar^2 k^2}{2m_{eq}} + E_{gap} \quad , \tag{61}$$

where $m_{el}$ and $m_{hole}$ are the effective mass of the electron and hole in the conduction and valence band, respectively, $m_{eq}$ is the equivalent mass for the electron-hole pair and $E_{gap}$ is the band gap energy. Since the momentum of the photon is very small relative to that of the electron we assume that the transitions occur mostly between initial and final states with opposite vector states $\vec{k}$ and $-\vec{k}$. We notice that the energy $\varepsilon(k)$ is a function of $k$ and also of the band gap $E_{gap}$. The Coulomb attraction between electron and hole in a single pair can be taken into account in (61) by lowering somewhat $E_{gap}$ effective value (see e.g. [25]).

In order to relate (61) to a quantum mechanical Hamiltonians we have to take into account that the electron or hole satisfy separately the fermion anti-commutation relations. The special point in the present discussion is that we treat in the plane electrons-holes pairs. In the previous Sections we described a variation of the Dirac equation described in the plane but the anti-commutation relations for the electrons or holes is an intrinsic property (see e.g. [12-13]) which is not changed by the restriction of the dynamics in a plane. We should notice also that in analogous way to the analysis given in (54-56) only the difference in energies is important and not their absolute values so that (61) expresses the difference in energy between the conduction and valence electrons. Although we use the similarity between the relativistic electron-hole theory and the electrons-holes pairs model we should take into account that the band gap energy $E_{gap}$ is not relativistic and might be very small. Also the electron or hole mass is changed to $m_{el}$ or $m_{hole}$, respectively, due to solid state effects, with the mass $m_{eq}$ for the electron-hole pair. Orders of magnitudes, for example, for gallium arsenide (GAAS):

$E_{gap} \approx 1.5 eV \quad ; \quad m_{hole} \approx 0.1 m_{electron} \quad ; \quad m_{el} \approx 0.065 m_{electron}$ .

Our main interest is, however, in high-TC materials which might have quite different parameters. A possible quantum representation for a certain hole-particle production and annihilation operator is given as:

$$\hat{O}_{hole-elec}(k_x, k_y) = \left[\hat{b}^\dagger(k_x, k_y)\hat{d}^\dagger(-k_x, -k_y) + \hat{b}(k_x, k_y)\hat{d}(-k_x, -k_y)\right] \quad , \tag{62}$$

where $\hat{O}_{hole-elec}(k_x,k_y)$ represents the operator for producing and annihilating of electron-hole pair in the plane with wave vector components $k_x, k_y$ for electrons, and wave vectors components $-k_x, -k_y$ for holes.

One should notice the following characteristics of $\hat{O}_{hole-elec}(k_x,k_y)$:

a) In spite of the non-unitary representations which have been used, the operator $\hat{O}_{hole-elec}$ is Hermitian.

b) In (62) it is assumed that the electron and hole in the electron-hole pair have opposite values of momentum (see e.g. [25-26]). Such assumption is justified approximately since, for example, the photon producing the electron-hole pair or emitted by recombination has a negligible momentum relative to that of the electron and the hole.

c) The charge and momentum are preserved under the interaction $\hat{O}_{hole-elec}$ of the system.

d) All operators operating on the electrons commute with the operators operating on the holes.

We find here an interesting quantum effect. While the creation and annihilation operators of $\hat{b}$ and $\hat{d}$ operators satisfy the anti-commutation relations given by (51-52) the operator $\hat{O}_{hole-elec}$ satisfy approximately the boson commutation-relations:

$$\left[\hat{b}(k_x,k_y)\hat{d}(-k_x,-k_y), \hat{b}^\dagger(k'_x,k'_y)\hat{d}^\dagger(-k'_x,-k'_y)\right] \propto \delta(k_x - k'_x)\delta(k_y - k'_y) \quad . \tag{63}$$

In deriving (63) we have used the anti-commutation relations of (52), twice. We have to take into account all possible hole-electron coupling in momentum space so that (62) should be integrated over different directions. The operator representing the number of hole-electron pairs in such system is given as

$$\hat{N}_{hole-elec} = \sum_{k_x,k_y} \hat{b}^\dagger(k_x,k_y)\hat{d}^\dagger(-k_x,-k_y)\hat{b}(k_x,k_y)\hat{d}(-k_x,-k_y) \quad . \tag{64}$$

The interesting result in the above formalism is that the operator $\hat{N}_{hole-elec}$ is a boson operator, due to the commutation-relations (63) and there is no restriction on this number as 'Pauli exclusion principle' for separate electrons and holes is not valid here.

While there might be other different perturbations which will produce the electrons-holes pairs, the essential point is that annihilation and creation operators for such states satisfy boson commutation relations. For simplicity of calculations we assume that the intensity of the perturbation (62) depends on the absolute value of $\vec{k}$ and not on its direction, in agreement with (61). We will show by the following analysis that above a certain critical value for the electrons-holes-pairs surface density, many electrons-holes pairs are included within distances which are order of the pair wavelength, so that (62) does not describe the annihilation and creation operators for separate pairs but that corresponding to the quantum collective many pairs system. We will show that scattering effects between different pairs can lead to Bose-Einstein condensation. The effect is different from that of the BCS theory due to the properties of electrons-holes-pairs. Under external electric field *collective* amount of holes will move in the electric field direction and the corresponding collective amount of electrons will move in the opposite direction, so that effective strong current will be produced. In order to analyze this effect we need to consider first the electrons-holes momentum distribution for the case of negligible scattering effects, using Bose-Einstein statistics as follows:

For homogeneous distribution of electrons-holes pairs in the $(\vec{k}_x, \vec{k}_y)$ space, the mean occupation number of the boson state with energy $\varepsilon(k)$ is given by (see e.g. [14]):

$$\langle n_{\vec{k}} \rangle = \frac{1}{\exp\{\beta[\varepsilon(k) - \mu]\} - 1} \quad , \tag{65}$$

where $\mu$ is the chemical potential, $\varepsilon(k)$ is the single particle energy and $\beta = 1/k_B T$. The average number $\langle N \rangle$ for the total number of electrons-holes pairs in the planar system is given by

$$\langle N \rangle = \frac{A}{(2\pi)^2} \iint \frac{dk_x dk_y}{\exp\{\beta[\varepsilon(k) - \mu]\} - 1} \quad , \tag{66}$$

where $A$ is the area of the planar system. Eq. (66) for the planar surface is analogous to a similar equation in the three-dimensional space where there ([14]) we have the $\int d^3 k$. The approximation of replacing the sum over values of $\vec{k}$ by an integral is valid for the planar system for any temperature, as long as scattering effects between different pairs are neglected. The Bose condensation will be obtained in the planar system by scattering effects which will be treated in the next Section, while in the present Section such effects are neglected.

According to the above solid state model

$$\varepsilon(k) = \frac{\hbar^2 (k_x^2 + k_y^2)}{2m_{eq}} + E_{gap} \quad . \tag{67}$$

Using the assumption of homogenous distribution in the $(k_x, k_y)$ coordinates, we exchange the integration $\iint dk_x dk_y$ to $\int 2\pi k_r dk_r$ where $k_r = \sqrt{k_x^2 + k_y^2}$ and substitute (67) into (66). Then we get:

$$\frac{\langle N \rangle}{A} = \frac{1}{(2\pi)^2} \int dk_r \frac{2\pi k_r}{F \exp\{\beta[(\hbar^2 k_r^2 / 2m_{equ}) + E_{gap}]\} - 1} \quad ; \quad F = \exp(-\beta\mu) \quad . \tag{68}$$

Substituting $k_r^2 = x$, $dx = 2k_r dr$ in (68), we get:

$$\begin{aligned}\frac{\langle N \rangle}{A} &= \frac{1}{4\pi} \int dx \frac{1}{F \exp\{\beta[(\hbar^2 x / 2m_{equ}) + E_{gap}]\} - 1} \\ &= \frac{1}{4\pi} \int dx \frac{F^{-1} \exp\{-\beta[(\hbar^2 x / 2m_{equ}) + E_{gap}]\}}{1 - F^{-1} \exp\{-\beta[(\hbar^2 x / 2m_{equ}) + E_{gap}]\}}\end{aligned} \tag{69}$$

The critical point in the present calculation is that we get explicit result for the surface density of electrons-holes pairs but the integration should be taken into certain limits as follows:

$$\frac{\langle N \rangle}{A} = \frac{2m_{equ}}{4\pi\beta\hbar^2} \ln\left[1 - F^{-1} \exp\{-\beta[(\hbar^2 x / 2m_{equ}) + E_{gap}]\}\right]_{x=0}^{x=k_f^2} \quad . \tag{70}$$

Here we find that we have a certain band of electrons-holes pairs energies where in the lower limit we have $x = 0$, i.e., $\vec{k} = 0$ and in the upper limit $x = k_f^2$. Assuming that we have obtained experimentally a certain intensity of electrons-holes pairs given by $\langle N \rangle / A$, at temperature $T$, *then the value of the chemical potential $\mu$ is a function of this surface density.*

Under the condition that $[(\hbar^2 k_f^2 / 2m_{equ}) + E_{gap}]\beta \gg 1$ the upper limit in (70) can be neglected and then we get

$$\frac{\langle N \rangle}{A} = -\frac{2m_{equ}}{4\pi\beta\hbar^2} \ln\left[1 - F^{-1} \exp\{-\beta E_{gap}\}\right] \quad . \tag{71}$$

By calculating an appropriate value for $F$, (71) can give the experimental value for $\langle N \rangle / A$.

By using the anti-log in (71) we get:

$$\exp\left[-\frac{4\pi\beta\hbar^2}{2m_{equ}} \frac{\langle N \rangle}{A}\right] = \left[1 - F^{-1} \exp\{-\beta E_{gap}\}\right] \quad . \tag{72}$$

I find according to the above analysis that Bose-Einstein condensation is not obtained in the planar system under the condition that scattering effects can be neglected. We show in the next Section that scattering effects can lead to condensation into Bose ground state under the condition of high density electrons-holes pairs.

From (72) we get the relation

$$F \exp\{\beta E_{gap}\} = \frac{1}{1 - \exp\left[-\frac{4\pi\beta\hbar^2}{2m_{equ}} \frac{\langle N \rangle}{A}\right]} . \qquad (73)$$

Then, by substituting this result in (65) we get

$$\langle n_k \rangle \simeq \frac{1}{\exp\{\beta\hbar^2 k^2 / 2m_{equ}\}\left\{1 - \exp\left[-\frac{4\pi\beta\hbar^2}{2m_{equ}} \frac{\langle N \rangle}{A}\right]\right\}^{-1} - 1} . \qquad (74)$$

The number of electrons-holes pairs in the range between $k_r = \sqrt{k_x^2 + k_y^2}$ and $k_r + dk_r$ is given by multiplying $\langle n_k \rangle$ with $2\pi k_r dr$. In order to demonstrate the effect of the correction term included in the squared brackets of (74) let us put some numbers. Assuming, for example, $m_{equ} = 0.1 m_{electron}$, $T = 100^0 K$. Then

$$\exp\left[-\frac{4\pi\beta\hbar^2}{2m_{equ}} \frac{\langle N \rangle}{A}\right] \approx \exp\left[-5.6 \cdot 10^{-12} \langle N \rangle / A\right] , \qquad (75)$$

where $A$ is given in units of $cm^2$. Under the condition $\frac{\langle N \rangle}{A} \gg 5 \cdot 10^{11}$ the expression (75) becomes quite small and under this condition

$$\langle n_k \rangle \simeq \frac{1}{\exp\{\beta\hbar^2 k^2 / 2m_{equ}\} - 1} , \quad for \quad \{\beta\hbar^2 k^2 / 2m_{equ}\} > 1, \qquad (76)$$

Then, most electrons-holes pairs will have energies of order of magnitude $k_B T$. Under the condition $(4\pi\beta\hbar^2 / 2m_{equ})(\langle N \rangle / A) \ll 1$ the distribution will be much broader. One should take into account that for values of $k$ tending to zero, $\langle n_k \rangle$ of (76) diverges and the divergence is eliminated only by using the more accurate expression (74).

There is a fundamental difference between cooperative effects in a planar system and those of a three-dimensional system. The average distance between $N$ boson particles in a three dimensional system is given by $(V / N)^{1/3}$, where $V$ is the volume, while for a planar systems it will become smaller as is given by $(A / N)^{1/2}$ where $A$ is the area of the system. So that attractive forces between different pairs will be more important for planar systems. By assuming in an example $\left[A/\langle N\rangle\right]^{1/2} \approx \left(1/10^{12}\right)^{1/2} = 10^{-6} cm$, then the average distance between neighboring electrons-holes pairs will be of order $100 A^0$.

The average wavelength of the electron-hole pair ('exciton'), under the conditions of (76), can be estimated as follows:

$$\lambda_{eff} = \frac{h}{m_{equ} v_{exc}} \approx \frac{h}{\sqrt{m_{equ} kT}} \quad ; \quad v_{exc} \approx \frac{kT}{m_{equ}} \quad . \tag{77}$$

Assuming, for example, $m_{equ} = 0.1 m_{electron}$, $T = 100^0 K$, then we get $\lambda_{eff} \approx 6 \cdot 10^{-5} cm$. We find for $\frac{\langle N\rangle}{A} = 10^{12}$ average distance $d \simeq 10^{-6} cm$, so that $\lambda_{eff}$ is greater by a factor 60 than the average neighboring distance between electrons-holes pairs, and approximately $60^2 = 3600$ electrons-holes pairs are included within an area with dimensions of electron-hole pair wavelength. In conclusion, for higher surface density of electrons-holes pairs, collective distribution will be obtained and the analysis becomes quite different as analyzed in the next Section leading to Bose-Einstein condensation.

## 5  Electrons-holes-pairs with high density leading to Bose condensation

We use a short notation where the electron-hole-pair is given by $\left|\vec{k}_{elec}, -\vec{k}_{hole}\right\rangle$, and where $\vec{k}_{elec} = (k_x, k_y)_{elec}$ and $-\vec{k}_{hole} = \left(-k_x, -k_y\right)_{hole}$, describing these vectors in the plane. For high surface density of electrons-holes pairs, where many pairs are included in an area of wavelength dimensions, attractive interaction is produced between such electrons-holes pairs. Such single interaction can be described by an attractive potential between the pair $\left|\vec{k}_{elec}, -\vec{k}_{hole}\right\rangle$ and $\left|\vec{k}'_{elec}, -\vec{k}'_{hole}\right\rangle$.

If $w_k$ is the probability that the pair $\left|\vec{k}_{elec}, -\vec{k}_{hole}\right\rangle$ is produced, then the energy of such electron-hole-pair is given by

$$E_{\vec{k}} = w_{\vec{k}} \varepsilon_k = w_k \left(\frac{\hbar^2 k^2}{2 m_{equ}} + E_{gap}\right) \quad , \tag{78}$$

corresponding the result given in (61). Since a pair state $\left|\vec{k}_{elec}, -\vec{k}_{hole}\right\rangle$ can be either occupied or unoccupied, we choose a representation of two orthogonal states $\left|1\right\rangle_{\vec{k}}$ and $\left|0\right\rangle_{\vec{k}}$ where $\left|1\right\rangle_{\vec{k}}$ is the state in which $\left|\vec{k}_{elec}, -\vec{k}_{hole}\right\rangle$ is occupied and $\left|0\right\rangle_{\vec{k}}$ is the corresponding unoccupied state. The general state of the pair $\left|\vec{k}_{elec}, -\vec{k}_{hole}\right\rangle$ is given by

$$\left|\psi\right\rangle_{\vec{k}} = u_{\vec{k}}\left|0\right\rangle_{\vec{k}} + v_{\vec{k}}\left|1\right\rangle_{\vec{k}} \quad ; \quad w_{\vec{k}} = v_{\vec{k}}^2 \quad ; \quad 1 - w_{\vec{k}} = u_{\vec{k}}^2 \quad , \tag{79}$$

where for simplicity of calculations we assumed that the amplitudes $u_{\vec{k}}$ and $v_{\vec{k}}$ are real. The ground state of the many-body system can be described by the product

$$\left|\Phi\right\rangle_G = \prod_{\vec{k}}\left(u_{\vec{k}}\left|0\right\rangle_{\vec{k}} + v_{\vec{k}}\left|1\right\rangle_{\vec{k}}\right) \quad . \tag{80}$$

In this approximation the many-body state can be described in terms of non-interacting electrons-holes pairs. In two-dimensional representation

$$\left|1\right\rangle_{\vec{k}} = \begin{pmatrix} 1 \\ 0 \end{pmatrix}_{\vec{k}} \quad , \quad \left|0\right\rangle_{\vec{k}} = \begin{pmatrix} 0 \\ 1 \end{pmatrix}_{\vec{k}} \quad . \tag{81}$$

The creation and annihilation operators for electrons-holes pairs are described, respectively, by

$$\sigma_{\vec{k}}^+ = \begin{pmatrix} 0 & 1 \\ 0 & 0 \end{pmatrix} \quad ; \quad \sigma_{\vec{k}}^- = \begin{pmatrix} 0 & 0 \\ 1 & 0 \end{pmatrix} \quad . \tag{82}$$

The total Hamiltonian of interactions between electron-hole-pairs $\left|\vec{k}_{elec}, -\vec{k}_{hole}\right\rangle$ and electron-holes-pairs $\left|\vec{k}'_{elec}, -\vec{k}'_{hole}\right\rangle$ is given by summation over all pairs interactions as (see, e.g. [24]):

$$\mathrm{H} = -\frac{1}{L^2} V_0 \sum_{\vec{k},\vec{k}'} \sigma_{\vec{k}}^+ \sigma_{\vec{k}'}^- \quad . \tag{83}$$

We assumed here an average attractive potential $V_0$ between electrons-holes pairs and $L^2$ represents the area of quantization (analogous to the volume quantization $L^3$ in the three-dimensional case). Following our previous description $L$ is of order of the electron-hole pair wavelength. A straight-forward calculation shows that the total attractive energy (see e.g. [24]) is given by

$$\left\langle\Phi\left|\mathrm{H}\right|\Phi\right\rangle_G = -\frac{V_0}{L^2} \sum_{\vec{k},\vec{k}'} v_{\vec{k}} u_{\vec{k}} u_{\vec{k}'} v_{\vec{k}'} \quad . \tag{84}$$

Adding the energy of non-interacting pairs given by (78) to the attractive energy of (84) the total energy of the ground state is given by

$$W_G = \sum_{\vec{k}} v_{\vec{k}}^2 \varepsilon_{\vec{k}} - \frac{V_0}{L^2} \sum_{\vec{k},\vec{k}'} v_{\vec{k}} u_{\vec{k}} u_{\vec{k}'} v_{\vec{k}'} \quad . \tag{85}$$

We define

$$v_{\vec{k}} = \cos\theta_{\vec{k}} \quad ; \quad u_{\vec{k}} = \sin\theta_{\vec{k}} \quad . \tag{86}$$

Then (85) can be written as

$$W_G = \sum_{\vec{k}} \varepsilon_{\vec{k}} \cos^2\theta_{\vec{k}} - \frac{1}{4}\frac{V_0}{L^2} \sum_{\vec{k},\vec{k}'} \sin(2\theta_{\vec{k}})\sin(2\theta_{\vec{k}'}) \quad . \tag{87}$$

For the minimum of $W_G$ we get

$$\frac{\partial W_G}{\partial \theta_{\vec{k}}} = -\varepsilon_{\vec{k}} \sin(2\theta_{\vec{k}}) - \frac{V_0}{2L^2}\cos(2\theta_{\vec{k}})\sum_{\vec{k}'}\sin(2\theta_{\vec{k}'}) = 0 \quad . \tag{88}$$

We define:

$$\Delta = \frac{V_0}{2L^2}\sum_{\vec{k}'}\sin(2\theta_{\vec{k}'}) = \frac{V_0}{L^2}\sum_{\vec{k}'} u_{\vec{k}'} v_{\vec{k}'} \quad ; \quad E_{\vec{k}} = \sqrt{\varepsilon_{\vec{k}}^2 + \Delta^2} \tag{89}$$

Then we get

$$\Delta = -\varepsilon_{\vec{k}} \tan(2\theta_{\vec{k}}) \quad ; \quad E_{\vec{k}}^2 = \varepsilon_{\vec{k}}^2 + \Delta^2 = \Delta^2(1+\cot^2(2\theta_{\vec{k}})) \to \Delta = E_{\vec{k}} \sin(2\theta_{\vec{k}}) \quad , \tag{90}$$

$$\sin(2\theta_{\vec{k}}) = 2u_{\vec{k}} v_{\vec{k}} = \frac{\Delta}{E_{\vec{k}}} \quad ; \quad \sum_{\vec{k}'}\sin(2\theta_{\vec{k}'}) = \sum_{\vec{k}}\sin(2\theta_{\vec{k}}) = \sum_{\vec{k}}\frac{\Delta}{E_{\vec{k}}} \quad . \tag{91}$$

In (91) the summation over $\vec{k}'$ is exchanged into summation over $\vec{k}$ since both are equivalent variables. We substitute (91) in (89) and then we get

$$\Delta = \frac{V_0}{2L^2}\sum_{\vec{k}}\frac{\Delta}{E_{\vec{k}}} = \frac{V_0}{2L^2}\sum_{\vec{k}}\frac{\Delta}{\sqrt{\varepsilon_{\vec{k}}^2 + \Delta^2}} \quad . \tag{92}$$

Consistency demands that $\Delta$ can be calculated by

$$1 = \frac{1}{2}\frac{V_0}{L^2}\sum_{\vec{k}}\frac{1}{\sqrt{\varepsilon_{\vec{k}}^2 + \Delta^2}} \quad . \tag{93}$$

Following the above analysis $\Delta$ is the condensation gap which can be calculated by (93) (see e.g. [24]). In order to take into account the summation over all states which have wave vector $\vec{k}$ we replace the summation by integral as

$$L^{-2}\sum_{\vec{k}} \to (1/2\pi)^2 \int d\vec{k} = (1/2\pi)^2 \int 2\pi k\, dk \quad ; \quad k = \sqrt{k_x^2 + k_y^2} \quad . \tag{94}$$

Also we multiply $\varepsilon_{\vec{k}}$ by $\langle n_{\vec{k}} \rangle$ , i.e., by the number of pair states with a wave vector $\vec{k}$.

Then the self-consistency equation becomes

$$1 = \frac{V_0}{4\pi} \int k dk \left\{ \left[ \varepsilon_k \langle n_k \rangle \right]^2 + \Delta^2 \right\}^{-1/2} \quad . \tag{95}$$

Substituting $\varepsilon_k$ from (61) and $\langle n_k \rangle$ from (74) and changing the variables by $k^2 = x$ ; $dx = 2kdk$ then the self-consistency equation becomes

$$1 = \frac{V_0}{8\pi} \int dx \left\{ \left( \hbar^2 x / 2m_{equ} + E_{gap} \right)^2 \left( \frac{\exp\left(\beta \hbar^2 x / 2m_{equ}\right)}{1-\alpha} - 1 \right)^{-2} + \Delta^2 \right\}^{-1/2} \quad . \tag{96}$$

According to (74) we defined here

$$\alpha = \exp\left[ -\frac{4\pi \beta \hbar^2}{2m_{equ}} \frac{\langle N \rangle}{A} \right] \quad . \tag{97}$$

Eq. (96) can be solved numerically for obtaining the condensation gap $\Delta$. One should notice that the temperature enters in (96), and in the correction term which also is temperature dependent. The calculation for $\Delta$ by (96) depends on four empirical constants: $V_0, m_{equ}, E_{gap}$, and $\frac{\langle N \rangle}{A}$.

Eq. (96) can be simplified by using the following change of variables

$$\hbar^2 x / 2m_{equ} = E \quad ; \quad \left( \beta \hbar^2 x / 2m_{equ} \right) = \frac{E}{kT} \quad ; \quad dx = \frac{2m_{equ}}{\hbar^2} dE \quad . \tag{98}$$

Then (96) can be written as

$$1 = \frac{V_0}{8\pi} \frac{2m_{eq}}{\hbar^2} \int dE \left\{ \left( E + E_{gap} \right)^2 \left( \frac{\exp(E/kT)}{1-\alpha} - 1 \right)^{-2} + \Delta^2 \right\}^{-1/2} \quad . \tag{99}$$

It is convenient to obtain $\Delta$ in $kT$ units of energy.

## 6 Electrons-holes-pairs condensation and the chemistry of high-TC materials

It has been shown in previous Section that under the condition that many electrons-holes pairs are produced within an area of the electron-hole-pair wavelength dimensions, a condensation to a low ground state can occur. While the condensation to this ground state is similar in some respects to the BCS theory, we notice that the present model for electrons-holes-pairs condensation is different from that of the BCS theory where there the condensation is made by electrons pairs with an interaction mediated by phonon interactions. In the present model the Bose condensation has been related to electrons-hole pairs and the analysis has been made for planar systems depending on quantum field theory for electrons-holes pairs.

I claim that there are close relations between the present electrons-holes-pairs model and the complicated structures of high TC materials. I find that in these materials electrons-holes pairs are produced which are usually referred to as 'excitons', and only the mechanism can be different from one system to another (see e.g. [1-6,27-31]). An example of electron-hole couple, with Coulomb interaction can be given by [27]:

$$-E_{elec-hole} = \frac{1}{4\pi\varepsilon_0} \frac{e^2}{d_{cu-o}} \qquad . \qquad (100)$$

Here holes and electrons correspond to $O^-$ and $Cu^+$, respectively. There are much more complicated charge transfer mechanisms [27-31] but from the point of phase transition of Bose condensation (see e.g. [32]) the analysis by quantum field theory will be similar.

## 7  Summary, discussion and conclusions

It has been shown that the Dirac equation, can be varied to a Dirac equation in the plane. Thus we are using a reduced Dirac equation of the form of (1), in the low dimensional space, where the perpendicular coordinate z is eliminated, the Pauli matrices are exchanging the $\gamma$ matrices, and the present wave functions are *two-dimensional*. These wave functions are given as the wave function $\Psi_+$ of (16) with the parameter $G_1(k_x,k_y)$ calculated in (24), and the wave function $\Psi_-$ of (26) with the parameter $G_2(k_x,k_y)$ calculated in (32). Since we use finite non-unitary representations, in analogous way to that of the Dirac equation, we insert a certain metric by which the normalization relations are calculated and given in (38-39). We get positive and negative solutions.

In Section 3, in analogy to the ordinary Dirac equation, field theory is presented where the wave functions are operators expanded into linear combinations of plane wave solutions with creation and annihilation operators satisfying the anti-commutation relations of (51-52). In this theory the creation operators $d^\dagger$ for holes are combined with annihilation operators $\hat{b}$ for electrons, and vice versa. The Hamiltonian and momentum are obtained, respectively, in (56) and (58). While the energy of the holes seems to be negative the relevant effective energy is the difference of energy of between electrons and holes so that we have to add their absolute energies. Such approach is obtained by a rigorous quantum mechanical treatment in which the Hamiltonian of (54) is transformed to that of (56).

I claim in the present work that the relativistic equation which describes electron-hole pairs can be applied with certain modifications to electron-holes pairing in solid state physics. In Section 4, I follow his idea, using a model for electron-hole pairing which describes the transition of electron from the valence band to conduction band. The energy of such electron-hole pair is described approximately by (61). In this equation the relativistic energy $2mc^2$ is exchanged into $E_{gap}$, where the energy $E_{gap}$ is not relativistic and it is rather very small. In the calculations for the momenta of the electrons we take into account an effective mass for the conduction electron given as $m_{el}$ and for the valence electron given as $m_{hole}$, which are different from the ordinary electron mass $m_{electron}$ due to solid state physics effects. A perturbation Hamiltonian $\hat{O}_{hole-elec}(k_x, k_y)$ for the production of electron-holes pairs is described in (62), and various properties of such perturbation, are analyzed. It is shown that while the creation and annihilation operators of separate electrons or holes satisfy the anti-commutation relations, the operator for producing and annihilating of electron-hole pair satisfy the boson commutation relations given by (63). We have calculated the Einstein Bose distribution for electrons-holes -pairs and found the conditions under which many electrons-holes are included in an area with wavelength dimensions.

In Section (5) we have treated scattering effects in high density of electrons-holes pairs leading to Bose condensation. The mechanism of Bose condensation by electrons-holes pairs is different from the BCS theory as the attractive forces between different pairs is based on boson exchange forces taken into account in (84) with an empirical attractive constant $V$ introduced in (83). The Bose condensation depends on the production of energy gap $\Delta$ which has been calculated for the present system in (96) or in a more simple form in (99). In Section 6 we have presented some comments on the relation between the present model and the complicated structure of high TC materials.